\documentclass[nofootinbib,twocolumn,showpacs,showkeys]{revtex4}
\usepackage{graphicx,color}
\usepackage{amssymb}
\usepackage{amsmath}
\usepackage{bm}
\usepackage[normalem]{ulem}
\usepackage{hyperref}

\begin{document}

\title{Nontrivial static, spherically symmetric vacuum solution in a nonconservative theory of gravity }

\author{A. M. Oliveira$^{1,2}$}\email{adriano.oliveira@ifes.edu.br}
\author{H. E. S. Velten$^{2}$}\email{velten@pq.cnpq.br}
\author{J. C. Fabris$^2$ }\email{fabris@pq.cnpq.br}

\affiliation{$^1$Instituto Federal do Esp\'irito Santo (IFES), Guarapari, Brazil}
\affiliation{$^2$Universidade Federal do Esp\'{\i}rito Santo (UFES), Vit\'oria, Brazil}

\begin{abstract}
We analyse the vacuum static spherically symmetric space-time for a specific class of non-conservative theories of gravity based
on the Rastall's theory. We obtain a new vacuum solution which has the same structure as the Schwarzschild-de Sitter solution in the General Relativity theory obtained with a cosmological constant playing the r\^ole of source. 
We further discuss the structure (in particular, the coupling to matter fields) and some cosmological aspects of the underline non-conservative theory.

\keywords{Gravity; General Relativity; Rastall gravity; Black holes}
\pacs{04.50.Kd; }
\end{abstract}

\maketitle

\section{Introduction} 

Most of the alternative theories to General Relativity (GR) have been designed in the context of the dark matter/energy phenomena. This approach avoids the introduction of extra dark fields by promoting a change in the structure of how the gravitational interaction works. Among them (see \cite{MG} for references on mofidied gravity theories), a few take a major departure from one of the main cornestones of GR i.e., the conservation laws expressed by the null divergence of the energy-momentum tensor $T^{\mu\nu}$. The Rastall gravity is one example of such proposal \cite{rastall} where the divergence of $T^{\mu\nu}$ is proportional to the gradient of the Ricci scalar, such that the usual conservation laws are recovered in flat space-time. Since the conservation of $T^{\mu\nu}$ is a consequence of the minimal coupling of the matter fields to the gravity sector and the invariance by diffeomorphism, at least one of these conditions must be suppressed in order to provide to Rastall gravity theory a Lagrangian formulation. Indeed, this proposal still suffers with the lack of a consistent Lagrangian structure and therefore the Rastall gravity remains, for the moment, a phenomenological modification of the GR theory. This fact does not impede the investigation of possible solutions for the Rastall's field equations which is one of the aims of this work.   

The field equations in this non-conservative theory are,
\begin{eqnarray}
\label{eq1}
R_{\mu\nu} - \frac{\lambda}{2}g_{\mu\nu}R &=& 8\pi G T_{\mu\nu},\\
\label{eq2}
{T^{\mu\nu}}_{;\mu} &=& \frac{1 - \lambda}{16\pi G}R^{;\nu}.
\end{eqnarray}
When $\lambda = 1$ the usual GR theory is recovered, with the usual conservation of the energy-momentum tensor.
The original motivation for such proposal is related to the fact that any conservation law is essentially tested in flat space-time (where $R = 0$). Therefore, in curved space-time some modifications of the conservation laws can occur. Moreover, quantum effects in curved space-times lead to a modification of the classical expressions for the energy-momentum tensor \cite{birrell}. 

The Rastall gravity has been tested in many different context (see for example Refs. \cite{daouda,velten}). In the cosmological domain it reproduces the success of the $\Lambda$CDM model on large scales whereas the small scale behavior can be different, exactly in the domain where the matter sector of the $\Lambda$CDM model faces some problems related to clustering and density distribtuion \cite{escala}.

It is worth noting that the nature of the Rastall's proposal is associated to high curvature environments and therefore astrophysical objects/configurations should provide a firm ground to testing it. Indeed the best constraints on Rastall gravity have been obtained via the study of neutron stars equilibrium configurations \cite{velten}. Therefore, further investigation static and spherically symmetric solutions of Rastall gravity can reveal new features of this theory. 

Following the reasoning above, black hole configurations can potentialy represent an interesting route of investigation for the Rastall gravity. In order to address this issue, let us look at vacuum solutions of equations (\ref{eq1}) and (\ref{eq2}). In this case, it is worth noting that these equations satisfy the following conditions:
\begin{eqnarray}
R_{\mu\nu} = 0, \quad R = 0.
\end{eqnarray}

Hence, the Schwarzschild solution is an expected solution of the Rastall's equation in vacuum. However, the large majority of modified gravity theories possess instead the Schwarzschild-de Sitter (SdS) configuration \cite{Iorio2016}. The SdS solution is the typical one when the cosmological constant $\Lambda$ is incorporated to the standard general relativity. In the case of modified theories, the effective cosmological constant is usually written in terms of parameters of the modified theory. Therefore, following the above reasoning, there is an apparent contradiction with the Rastall theory which, on cosmological scales, can be seen as a realization of the $\Lambda$CDM model \cite{daouda}.

The goal of the present paper is to show that in the realm of the Rastall theory, for the particular case where $\lambda = 1/2$, there is another---unexpected---vacuum static spherically solution which resembles the Schwarzschild-de Sitter solution. In that particular case there is no need to have a cosmological constant added to the basic formulation of the Rastall theory in order to generate such solution.

\section{The (new) non-trivial solution} 

We point out now the particular case of the Rastall's gravity which admits a well behaved static, spherically symmetric solution. We provide in the following a detailed (step-by-step) derivation of it. In order to construct such solution, let us consider equation (\ref{eq1}) in vacuum:
\begin{eqnarray}
R_{\mu\nu} = \frac{\lambda}{2}g_{\mu\nu} R.
\label{eq:CS_2189}
\end{eqnarray}
The trace of this equation reads,
\begin{eqnarray}
 R(1-2\lambda)=0.
\label{eq:RasR}
\end{eqnarray}
The above condition (\ref{eq:RasR}) is satisfied: (i) imposing $R=0$, implying that $R_{\mu\nu}=0$. Then, from (\ref{eq:CS_2189}), one recovers the usual Schwarzschild solution; or (ii) when $\lambda=1/2$. In this case, if one looks at equations (\ref{eq1}) and (\ref{eq2}) assuming the presence of matter, the latter condition is realised only if the trace of the energy-momentum tensor is zero (radiative fluid, for example). However, there are possible generalisations which will be discussed later. 

The vacuum case of (\ref{eq1}) with $\lambda = 1/2$ reads,
\begin{eqnarray}\label{VEeqlambda12}
R_{\mu\nu} = \frac{g_{\mu\nu}}{4} R.
\end{eqnarray}

The static, spherically symmetric metric is given by,
\begin{eqnarray}
\label{metric}
ds^2 = B(r)dt^2 - A(r)dr^2 - r^2\,d\Omega^2,
\end{eqnarray}
where $d\Omega^2$ represents the angular part of the metric.
Therefore, the components of the energy-momentum tensor read:
\begin{eqnarray}
R_{tt} &=&  \frac{B}{A}\biggr\{\frac{1}{2}\frac{B''}{B} - \frac{1}{4}\frac{B'}{B}\biggl(\frac{B'}{B} + \frac{A'}{A}\biggr) + \frac{1}{r}\frac{B'}{B}\biggl\},\\
R_{rr} &=& - \biggr\{\frac{1}{2}\frac{B''}{B} - \frac{1}{4}\frac{B'}{B}\biggl(\frac{B'}{B}+\frac{A'}{A}\biggr) - \frac{1}{r}\frac{A'}{A}\biggl\},\\
R_{\theta \theta} &=& \biggr\{A-1-\frac{r}{2}\biggr(\frac{B'}{B} - \frac{A'}{A}\biggl)\biggl\}\frac{1}{A}.
\end{eqnarray}

Applying these expressions to (\ref{VEeqlambda12}) we find 
\begin{eqnarray}
\frac{R_{tt}}{B}= \frac{R}{4} \hspace{0.3cm}{\rm and}\hspace{0.3cm} \frac{R_{rr}}{A} = -\frac{R}{4},
\end{eqnarray} 
which trivially results in,
\begin{eqnarray}
\left(\frac{B'}{B}+\frac{A'}{A} \right)=0.
\label{eq:Rastall_CS2193}
\end{eqnarray}
This relation is also present in the standard Schwarzschild solution when coordinates (\ref{metric}) are employed. Hence, for the two possibilities evoked above, $R = 0$ or $\lambda =1/2$, the usual relation between metric coefficients stands:
\begin{eqnarray}
B=A^{-1}.
\label{eq:Rastall_CS2194}
\end{eqnarray}

Let us consider the case $R_{\theta\theta} = 0$. Since $R$ and $R_{\mu\nu}$ vanish, we can determine the metric  
\begin{eqnarray}
B = 1+\frac{C_1}{r},
\end{eqnarray}
with $C_1= - 2 G \mathcal{M}$. The integration constant has been expressed in unities of $c^2$ and $\mathcal{M}$ plays the r\^ole of mass and $G$ being the Newtonian gravitational constant.

Now, looking at the second possibility in which $\lambda=1/2$, the $\theta-\theta$ component of Eq. (\ref{VEeqlambda12}) can be written as
\begin{eqnarray}
&1&+ \frac{r}{2A}\biggr\{\frac{A'}{A}-\frac{B'}{B}\biggl\}-\frac{1}{A} =-\frac{r^2}{4}\biggr[ \frac{B''}{AB} -\frac{B'}{2AB}\biggl(\frac{A'}{A}+\frac{B'}{B}\biggl)\nonumber\\&-&\frac{2}{rA}\biggr(\frac{A'}{A}-\frac{B'}{B}\biggl)-\frac{2}{r^2}\biggr(1-\frac{1}{A}\biggr)\biggl].
\end{eqnarray}
Using now relation (\ref{eq:Rastall_CS2193}), the above equality can be simplified leading to,
\begin{eqnarray}
1-\frac{r}{A}\frac{B'}{B}-\frac{1}{A} = -\frac{r^2}{4}\left[ \frac{B''}{AB}+\frac{4}{rA}\frac{B'}{B}-\frac{2}{r^2}\left(1-\frac{1}{A}\right)\right].\nonumber
\end{eqnarray}
Finally, with the help of (\ref{eq:Rastall_CS2194}) we find
\begin{eqnarray}
1-rB'-B= -\frac{r^2}{4}\left[B''+\frac{4}{r}B'-\frac{2}{r^2}\left(1-B\right)\right].\nonumber
\end{eqnarray}
This expression can be rewritten as,
\begin{eqnarray}
\frac{r^2B''}{2}-B+1=0,\nonumber
\end{eqnarray}
which can be integrated resulting in
\begin{eqnarray}
B(r)=1+\frac{C_1}{r}+C_2 r^2,
\label{eq:RasCS196}
\end{eqnarray}
where $C_2$ is an integration constant to be determined.

The static, spherically symmetric metric for the Rastall gravity theory with $\lambda = 1/2$ is therefore
\begin{eqnarray}\label{eq:RasCS197}
ds^2 &=& \left(1-\frac{2G\mathcal{M}}{r}+ C_2 r^2 \right) dt^2 \nonumber \\ &-& \left(1-\frac{2G\mathcal{M}}{r}+ C_2 r^2 \right)^{-1} dr^2 - r^2d\Omega^2.
\end{eqnarray}

The above result is the main finding of this work. It does not represent a new solution since the structure of the metric coefficients in (\ref{eq:RasCS197}) coincides with the Schwarzschild-de Sitter (or Schwarzschild-Anti de Sitter) solution obtained in GR theory with a cosmological constant \cite{frolov}. But we have demonstrated a new way to obtaining it even in a context in which the cosmological constant is absent.

With the identification $C_2 = -\Lambda/3$ the metric takes the form
\begin{eqnarray}
\label{SdS}
ds^2 &=& \left(1 - \frac{2G\mathcal{M}}{r} - \frac{\Lambda}{3} r^2 \right) dt^2 \nonumber \\ &-& \left(1- \frac{2G\mathcal{M}}{r} - \frac{\Lambda}{3} r^2 \right)^{-1} dr^2- r^2d\Omega^2,
\end{eqnarray}
which represents the Schwarzschild-de Sitter metric for $\Lambda > 0$ ($C_2 < 0$) and the Schwarzschild-Anti de Sitter metric for $\Lambda < 0$ ($C_2 > 0$). A detailed description of the causal structure of the Schwarschild-de Sitter metric can be found in Ref. \cite{gasperin} while for the Schwarzschild-Anti de Sitter space-time in Ref. \cite{chile}.

At this point it is worth mentioning the relationship of the Rastall theory ($\lambda=1/2$) and its EdS-like solution with other approaches.

In Ref. \cite{mannheim1989exact} Mannheim \& Kazanas have obtained a similar solution in the context of a conformal theory, based on the Weyl tensor. The field equations used in this reference contains a combination of the Ricci tensor and Ricci scalar and their derivatives, being very different from the Rastall's case analysed here. The main common property between the two cases is the fact that in both situations the trace of the vacuum field equations is zero. The static, spherically symmetric solution in the conformal gravity differs from the solution found in the Rastall case due to the existence of a term proportional to $r$, reflecting the different general structure of these theories. Then, solution (\ref{eq:RasCS196}) can be seen as a special case of the conformal gravity case with vanishing linear term.

More interesting is the relationship of our results with $f(R)$ theories. In the Palatini formalism, i.e., by varying independently the action of your theory with respect to the metric $g$ and the connection $\Gamma$, we obtain the following vacuum equations

\begin{equation}
f^{\prime}(R) R_{(\mu\nu)}(\Gamma)-\frac{1}{2}f(R) g_{\mu\nu} = 0,
\end{equation}
\begin{equation}
\nabla_{\alpha}^{\Gamma}[ \sqrt{g} f^{\prime}(R) g^{\mu\nu}]  =0
\end{equation}
where $\nabla^{\Gamma}$ is the covariant derivation with respect to the connection $\Gamma$.
There is also the scalar equation
\begin{equation}
f^{\prime}(R) R -2 f(R)=0.
\end{equation}
It is worth noting that the trace of the above equation is an algebraic equation for $R$ with solutions of the type $R=const$ and is identically satisfied if $f(R) \propto R^2$. For the latter case, regarding that $f^{\prime} (R) \neq 0$ the field equation becomes
\begin{equation}
R_{\mu\nu} - \frac{1}{4}R g_{\mu\nu}=0.
\end{equation}
This corresponds exactly to the case of Rastall gravity with $\lambda=1/2$ and therefore, it admits the same EdS-like vacuum solution.

In principle the Birkhoff theorem is valid for the Rastall structure analysed here. In fact, considering now the metric functions also as time dependent, $A = A(r,t)$ and $B = B(r,t)$, the $t-r$ component of the field equations implies,
\begin{eqnarray} 
\frac{1}{r}\frac{\dot A}{A} = 0,
\end{eqnarray}
leading to $A = A(r)$, and the derivation of the Birkhoff theorem follows the same lines as in the GR case, see
Ref. \cite{wei}. However, in order to have a more complete analysis, perhaps it would be interesting to investigated the interior solution and the matching conditions with the vacuum exterior solution.

\section{Final Discussion} 

 We have shown in this work that the Rastall gravity also admits an static and spherically symmetric vacuum solution resembling the Schwarzschild-de Sitter (SdS) one. This result is given by Eq. (\ref{eq:RasCS197}) where the integration constant $C_2$ plays the role af an effective cosmological constant. The determination of $C_2$ occurs via the standard procedure to determining $\Lambda$ in the SdS metric, i.e., one could use Solar-System data as proposed in Refs. \cite{SdSSS1, SdSSS2} or study the deflection of light in strong gravitational lensing systems \cite{SdSLens1, SdSLens2, SdSLens3}.

The phenomenology related to the SdS solution (independently of its origin) has been discussed in many places in the literature, and some specific considerations can be found in Ref. \cite{SdSSS1} and references therein. It must be stressed that the effects of the cosmological constant term are irrelevant at level of the solar system. In fact, from the solution (\ref{SdS}), we can identify two contributions to the total potential, one coming from the central mass (the usual Schwarzschild term) and the other coming from the cosmological constant. If we compute the force on a test mass
due to each of this contribution, and using the cosmological estimations for $\Lambda$,
\begin{eqnarray}
H_0^2 \sim \frac{\Lambda}{3},
\end{eqnarray}
$H_0$ being the Hubble function measured today,
we find that at the edge of the solar system (a distance from the Sun of about $40\,au$), the ratio between the force due to the cosmological term and the one related to the central mass
is of the order of $10^{-16}$: the effects of the $\Lambda$ are locally irrelevant. Both forces become comparable at
a distance of the order of $150\, pc$ \cite{sereno}

Using equation (\ref{eq2}), and fixing $\lambda = 1/2$, the vacuum configuration implies $R = R_0$ where $R_0$ is a constant. It is worth noting that $R_0=0$ for the Schwarzschild solution whereas $R_0 \neq 0$ for the
Schwarzschild-de Sitter or Schwarzschild-Anti de Sitter solution. We have shown also that it is possible to prove the Birkhoff theorem in the theory studied here, even if some dynamical collapsing configurations must be still carefully studied.

\begin{figure}[t]
\includegraphics[width=0.44\textwidth]{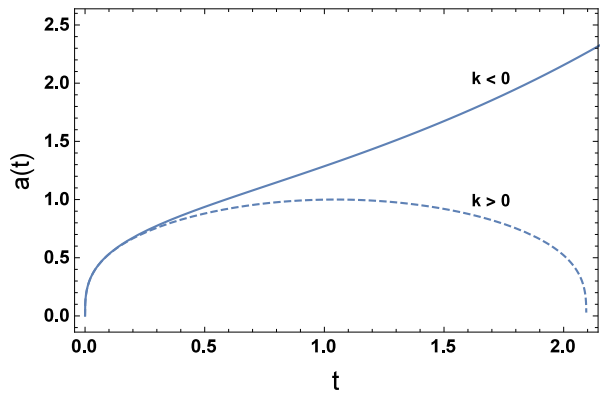}\\ 
\caption{Behaviour of the scale factor a(t). The constants were fixed to $a_0=1$, $n=3$ and $k=1$. }
\label{fig1}
\end{figure}

In choosing $\lambda = 1/2$, we end up with the field equation,
\begin{eqnarray}
\label{f-e}
R_{\mu\nu} - \frac{1}{4}g_{\mu\nu}R = 8\pi GT_{\mu\nu}.
\end{eqnarray}
Such coupling to matter is only possible for a vanishing trace of the energy momentum tensor ($T = 0$). This is the case for a radiative fluid which is conformally invariant. However, we can easily realise that the left-hand side of (\ref{f-e}) is not conformally invariant. This fact points out to the necessity of generalising the coupling to matter. For example, an equation sourced by an effective energy momentum-tensor $\mathcal{T}_{\mu\nu}$ such that
\begin{eqnarray}
R_{\mu\nu} - \frac{1}{4}g_{\mu\nu}R = 8\pi G \mathcal{T}_{\mu\nu}, \hspace{0.5cm}{\rm with}\hspace{0.5cm} \mathcal{T}=0,
\end{eqnarray}
would provide such compatibility. One simple possibility is,
\begin{eqnarray}
\label{g2}
R_{\mu\nu} - \frac{1}{4}g_{\mu\nu}R = 8\pi G\biggr\{T^c_{\mu\nu} + T_{\mu\nu}^m - \frac{1}{4}g_{\mu\nu}T^m\biggl\},
\end{eqnarray}
where the subscripts $c$ and $m$ designate the {\it conformal} and {\it matter} component respectively. This generalisation of the matter coupling to geometry, after a simple inspection, has some reasonable cosmological properties. Let us take, for example, equation (\ref{g2}) without the term $T^c_{\mu\nu}$. By employing a flat Friedmann-Lemaitre-Robertson-Walker metric to the above field equation is possible to verify that there is only one independent equation for the two variables, the scale factor $a$ and the fluid density $\rho$:
\begin{eqnarray}
\dot H = - 4\pi G(\rho + p),
\end{eqnarray}
where $H = \dot{a} / a$ is the Hubble function ($a$ is the scale factor), $\rho$ is the density and $p$ is the pressure. Hence, it is necessary to impose an ansatz for one variable in order to solve this equation. Assuming the equation of state $p = \omega\rho$, with $\omega$ constant, and imposing that the density scales as $\rho \propto a^{-n}$, $n$ being a number, there are, for any $\omega \neq - 1$ , two different types of scenarios: a closed-type expanding universe followed by a big crunch,
the scale factor being given by $a(t)=a_0 \sin^{2/n}(n \sqrt{k} t/2)$, where $a_0$ is the initial value of the scale factor and $k > 0$ is an integration constant; or an initial expanding universe whose behavior is determined by $n$ followed by a de Sitter phase, i.e., corresponding to the solution $a(t)=a_0 \sinh^{2/n}(n \sqrt{|k|}t/2)$, if $k<0$. The only case where no extra ansatz must be imposed corresponds to $p = - \rho$, leading to a de Sitter universe. These results can perhaps be connected with the presence of a cosmological constant in the static spherical symmetric case studied above. In order to illustrate such scenarios we show in Fig. 1 both solutions ($\omega \neq - 1$) for some representative values of the constants.

Deep investigations are necessary in order to verify the viability of such matter coupling proposal, as well as the possibility to obtain a action principle corresponding to this field equations. Our goal in the present work was to show how the Schwarzschild-de Sitter or the Schwarzschild-Anti de Sitter solutions emerge naturally, even without a cosmological constant as a source of the field equations, in a particular class of non-conservative theory of gravity.

\noindent
{\bf Acknowledgement:} We thank CNPq (Brazil) and FAPES (Brazil) for partial financial support.

\end{document}